\documentclass[aps,nofootinbib,superscriptaddress]{revtex4}

\usepackage{amsmath}
\usepackage{amssymb}
\usepackage{graphicx}
\usepackage{float}

\begin{document}

\title{Effective short-range interaction for spin-singlet $P$-wave nucleon-nucleon
scattering}
\author{Katie L. Ipson}
\affiliation{Theoretical Physics Group, School of Physics and Astronomy,
The University of Manchester, Manchester, M13 9PL, UK}
\author{Katharina Helmke}
\affiliation{Fakult\"at f\"ur Physik und Astronomie,
Ruhr-Universit\"at Bochum, D-44780 Bochum, Germany}
\author{Michael C. Birse}
\affiliation{Theoretical Physics Group, School of Physics and Astronomy,
The University of Manchester, Manchester, M13 9PL, UK}

\begin{abstract}

Distorted-wave methods are used to remove the effects of one-
and two-pion exchange up to order $Q^3$ from the empirical 
$^1P_1$ phase shift. The one divergence that arises can be 
renormalised using an order-$Q^2$ counterterm which is provided
by the (Weinberg) power counting appropriate to the effective field 
theory for this channel. The residual interaction is used to 
estimate the scale of the underlying physics.

\end{abstract}
\maketitle
\vspace{10pt}

Effective field theories (EFTs) provide important tools for constructing 
nuclear forces within a systematic framework.\footnote{For reviews, see 
Refs.~\cite{border,bvkrev,eprev}.} They respect the symmetries of underlying 
theory, in this case Quantum Chromodynamics, and offer model-independent 
descriptions of the dynamics at low energy scales. They rely on the existence of 
clear separation of scales between the low-energy physics of interest and
the underlying, high-energy (or short-distance) physics. This makes it possible 
to use a perturbation theory where the expansion parameters are ratios of 
low- to high-energy scales.

To construct such a theory one begins by writing down the most general Lagrangian 
or Hamiltonian for the relevant low-energy degrees of freedom, containing all
the interaction terms allowed by the symmetries of the system. The parameters
of the theory are the (initially unknown) constants multiplying these terms. 
There are in general an infinite number of these but they can be organised in 
terms of a ``power counting" of low-energy scales. To any given order in this 
counting, only a finite number of terms are needed.

Since Weinberg first suggested that these ideas could be applied to nuclear
forces \cite{weinberg}, there has been considerable debate about the 
appropriate power counting to use.\footnote{For recent summaries of two of 
these points of view, see Refs.~\cite{epelbaum} and \cite{birse09}.} Most 
of these arguments have been about channels where scattering is strong and 
some pieces of the potential may need to be treated nonperturbatively, 
namely the $S$ waves and the lower spin-triplet waves. In the higher partial 
waves, especially the spin-singlet ones, the scattering is weak and the 
picture is clearer. For these channels, Weinberg's original power counting 
is valid. This is based on simple dimensional analysis, counting powers of 
momenta and the pion mass. These low-energy scales are generically denoted 
by $Q$.

Many applications of EFTs test convergence by presenting theoretical 
nucleon-nucleon phase shifts, calculated at leading and higher orders, 
alongside empirical data. At each order, short-range interactions, expanded 
as polynomials in energy or momenta, have to be refitted to the data.
A powerful alternative is to use 
distorted-wave (DW) methods to remove the effects of known long-range 
forces from the observed phase shifts to leave residual short-range 
interactions. The sizes and energy-dependences of these can then be used 
to test the convergence of the EFT and to estimate the scales of the 
underlying physics. Such an approach was proposed in 
Ref.~\cite{bmcg} and applied to the peripheral spin-singlet waves. 
More recently it has been applied to spin-triplet waves \cite{birse07}
and the $^1S_0$ channel \cite{birse10} and closely related methods
have been used by other groups to study the $S$ waves \cite{shukla,pavon}.

In this note, we extend the method of Ref.~\cite{bmcg} by applying it to 
the $^1P_1$ channel. We first construct DWs for the leading-order 
one-pion-exchange (OPE) potential. Iterating this potential to all orders 
is not essential in a spin-singlet wave with $L\neq 0$ since Weinberg's 
power counting is expected to hold. However doing so avoids the need to
calculate terms to fourth order in perturbation theory. Since only the central 
part of OPE contributes and its $1/r$ singularity does not alter the power-law
behaviour of the wave functions near the origin, this treatment does not 
affect the power counting for the short-range pieces of the potential.

From the $K$ matrix that describes scattering between these DWs, we can
define a residual interaction. This still contains effects arising from
long-range potentials of orders $Q^2$ and above. We then use the
distorted-wave Born approximation (DWBA) to subtract the matrix elements
of the order-$Q^2$ and $Q^3$ two-pion-exchange (TPE) potentials
\cite{kbw,rtfs,friar}. This leaves a residual interaction that should 
represent purely short-range physics to this order.

One issue that arises here, but not for the channels studied in Ref.~\cite{bmcg},
is that the matrix elements of TPE between the DWs are divergent. This divergence
must be removed by renormalisation so that only finite quantities are being 
treated in perturbation theory. Weinberg's power counting provides one 
energy-independent counterterm, of order $Q^2$, in this channel and this is
sufficient to cancel the divergence. This is similar to what has been found in 
other waves where different counting schemes apply 
\cite{birse07,shukla,pavon,birse10}.\footnote{Note that here we follow the power
counting by treating the order-$Q^{2,3}$ potentials as perturbations. Other
approaches, which treat the whole potential to all orders, can lead to different
conclusions \cite{pvra,yep}.} 

We start by outlining the main features of the method from Ref.~\cite{bmcg}.
The DWs $\psi_{\scriptscriptstyle\rm OPE}(p,r)$ are obtained by solving the 
radial Schr\"odinger equation with the OPE and centrifugal potentials,
\begin{equation}
\frac{d^2 \psi_{\scriptscriptstyle\rm OPE}}{dr^2}
+\frac{2}{r}\,\frac{d\psi_{\scriptscriptstyle\rm OPE}}{dr}
-\left( \frac{L(L+1)}{r^2} + M_{\scriptscriptstyle N} 
V^{(0)}_{\scriptscriptstyle\rm OPE}(r)\right) 
\psi_{\scriptscriptstyle\rm OPE}(p,r)
=p^2\,\psi_{\scriptscriptstyle\rm OPE}(p,r).
\label{eq:RS}
\end{equation}
Here $L=1$ is the orbital angular momentum, $M_{\scriptscriptstyle N}$ is 
the nucleon mass and $p$ is the on-shell relative momentum in the 
centre-of-mass frame. We write the leading-order OPE potential in the same form 
as used in the Nijmegen analyses and take their preferred value for the $\pi N$ 
coupling, $f_{\pi\scriptscriptstyle NN}^2=0.075$ \cite{stoks1,stoks2}.

From the large-$r$ forms of these waves we can extract the OPE phase shift, 
$\delta_{\scriptscriptstyle\rm OPE}(p)$. Taking the difference between this and
the empirical phase shift, $\delta(p)$, we can define a residual $K$-matrix,
\begin{equation}
\tilde{K}(p) = - \frac{4 \pi}{M_{\scriptscriptstyle N}p}\, \tan \bigl( \delta (p) 
- \delta_ {\scriptscriptstyle\rm OPE}(p) \bigr).
\label{eq:Kmatrix}
\end{equation}
This describes the additional scattering between the DWs, produced by 
short-range interactions and long-range forces of order $Q^2$ and higher. 
If we take a $\delta$-shell form for the potential responsible
\begin{equation}\label{eq:vshort}
V_S(p,r)=\frac{9}{4\pi R_0^{4}}\,\tilde V(p)\,\delta(r-R_0),
\end{equation}
then we can extract its strength directly from $\tilde{K}(p)$, 
\begin{equation}
\tilde{V}^{(2)}(p) 
= \frac{R_0^{2}}{9\, \psi_{\scriptscriptstyle\rm OPE}(p,R_0)^2}
\, \tilde{K}(p),
\label{eq:V2}
\end{equation}
where the superscript (2) indicates that long-range effects of order $Q^2$ 
and higher are still present.

Finally, we use the DWBA to subtract the effects of long-range potentials to 
order $Q^3$. This leaves a residual short-range potential whose strength is
given by
\begin{equation}
\tilde{V}^{(4)}(p) 
= \frac{R_0^{2}}{ 9\, \psi_{\scriptscriptstyle\rm OPE}(p,R_0)^2} 
\left( \tilde{K}(p) - \langle \psi_{\scriptscriptstyle\rm OPE} (p) | 
V^{(2)}_{\scriptscriptstyle\rm OPE} + V^{(2,3)}_{\scriptscriptstyle\rm TPE} 
| \psi_{\scriptscriptstyle\rm OPE} (p) \rangle \right), 
\label{eq:V4}
\end{equation}
where the order-$Q^{2,3}$ TPE potentials can be found in Refs.~\cite{kbw,rtfs} 
and the corresponding order-$Q^2$ recoil correction to OPE is given by 
Friar \cite{friar}.

The TPE potential has terms with $1/r^6$ and $1/r^5$ singularities. 
Consequently, the radial integrals in its matrix elements between $P$ waves, 
which have the form $\psi(r)\propto r$ for small $r$, contain $1/r$ and $\ln r$ 
divergences. To rectify this, we regularise the integrals by imposing 
a radial cut-off at the same value of $R_0$ as used in the $\delta$-shell
potential. We can then renormalise the matrix elements by subtracting the sum 
of these divergent pieces (a constant which is independent of energy).

\begin{figure}[h]
\centering
\includegraphics[scale=0.9]{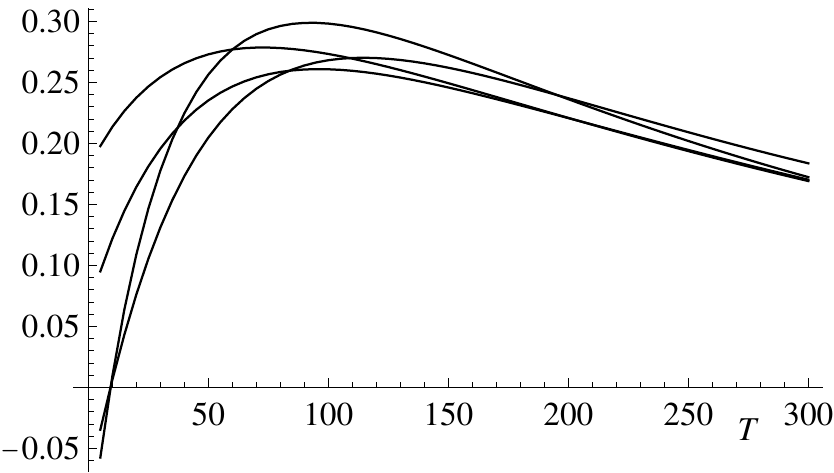}
\caption{The short-range effective potential $\tilde{V}^{(2)}(p)$, 
in fm$^{-4}$, plotted against the lab kinetic energy $T$ in MeV. 
The four curves correspond to different Nijmegen PWAs. A cut-off radius 
of $R_0=0.1$~fm was used.}
\label{fig:v2}
\end{figure}

We show first, in Fig.~\ref{fig:v2}, the effective short-range potential 
$\tilde{V}^{(2)}(p)$ in the $^1P_1$ partial wave after the removal of OPE only.
The cut-off radius of the $\delta$-shell was taken to be $R_0=0.1$~fm.
Results are shown for PWA93 and three different Nijmegen potentials 
\cite{stoks1,stoks2,NNonline}, all of which provide high-quality fits to the NN
scattering data. One can see that significant energy dependence is present 
at low energies, like that found in the higher partial waves \cite{bmcg}. This 
suggests that there may still be important contributions from long-range 
forces in this interaction. However we should note that the different Nijmegen 
analyses diverge for energies below about 50~MeV and so this region should not 
be regarded as well constrained by the data. This is because the centrifugal 
barrier prevents the nucleons from approaching each other closely at these 
energies and so the $^1P_1$ phase shift is completely dominated by OPE.

\begin{figure}[h]
\centering
\includegraphics[scale=0.9]{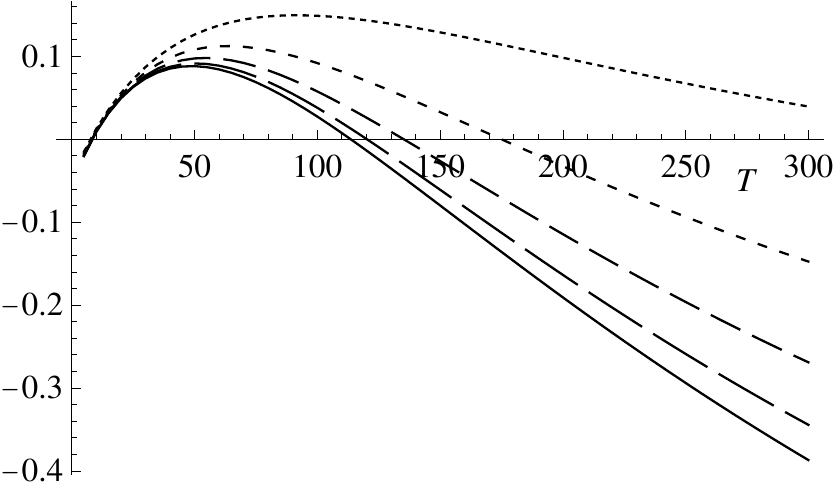}
\caption{The renormalised short-range effective potential $\tilde{V}^{(4)}(p)$,
in fm$^{-4}$, plotted against the lab kinetic energy $T$ in MeV. The different 
curves show the potentials obtained with cut-off radii $R_0=0.8$~fm 
(dotted), 0.4~fm, 0.2~fm, 0.1~fm, and 0.05~fm (solid).}
\label{fig:v4R0}
\end{figure}

\begin{figure}[h]
\centering
\includegraphics[scale=0.9]{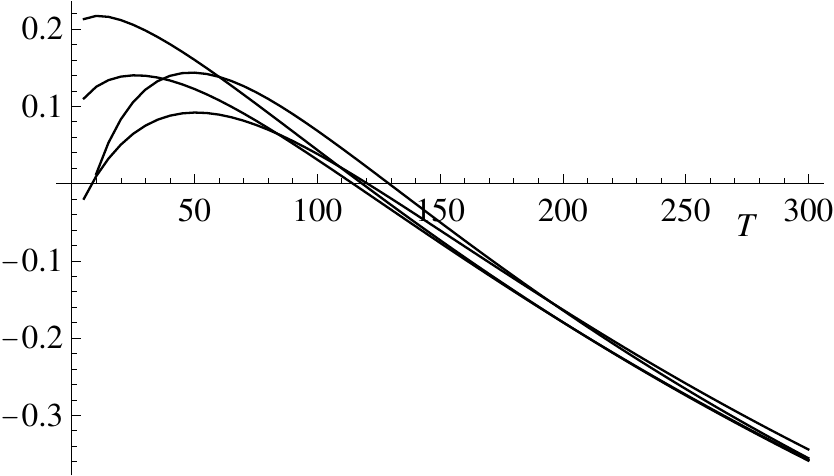}
\caption{The short-range effective potential $\tilde{V}^{(4)}(p)$ for four 
different Nijmegen PWAs. A cut-off radius of $R_0=0.1$~fm was used.}
\label{fig:v4}
\end{figure}

Using the DWBA to remove the effects of the order-$Q^{2,3}$ long-range potentials,
we obtain the results shown in Figs.~\ref{fig:v4R0} and \ref{fig:v4}. The matrix 
elements of the TPE potential have been renormalised by simply subtracting their 
values at some very low energy, $T=5$~MeV. 
Fig.~\ref{fig:v4R0} shows that the renormalised potential does indeed converge to a 
cut-off-independent form as $R_0\rightarrow 0$. For numerical convenience we present
our results for $R_0=0.1$~fm, for which any cut-off artefacts (terms in the potential
proportional to positive powers of $R_0$) are very small.

The resulting renormalised potentials $\tilde{V}^{(4)}(p)$ obtained from
several Nijmegen PWAs are shown in Fig.~\ref{fig:v4}. The ``turnover" at low
energies that was visible in Fig.~\ref{fig:v2} has been significantly reduced by
subtracting the higher-order long-range forces. However we should point out 
that much of this effect lies in the region where uncertainties in the PWAs 
make it hard to draw very strong conclusions.

To try to quantify the energy dependence of the residual potential, we made a 
least squares fit of a quadratic in the energy to $\tilde{V}^{(4)}(p)$. 
We fitted this over the energy range $T=100-200$~MeV, which was chosen to avoid 
both the low-energy region where the data are too inaccurate to determine 
the short-range potential reliably, and the high-energy region where our EFT is 
expected to converge slowly if at all. A quadratic was chosen because attempts 
to fit higher order polynomials to this range did not lead to stable values for 
the coefficients.

Writing the polynomial in the form
\begin{equation}
\tilde{V}^{(4)}(p) = a_0 + a_1 p^2 + a_2 p^4,
\end{equation}
we obtain $a_0 = 0.28$~fm$^4$, $a_1 = -0.20$~fm$^6$ and $a_2 = 0.0056$~fm$^8$.
This is dominated by a linear dependence on the energy ($p^2$), consistent with 
the visual impression of Fig.~\ref{fig:v4}. Without knowing the uncertainties on 
the phase shifts that are used to determine the potential, it is difficult to 
assign definite errors to these coefficients. However simple estimates suggest
that the values for the first two are reasonably accurate but $a_2$ is not.

For present purposes, our main interest in these coefficients is to estimate 
the scale of the short-distance physics represented by $\tilde{V}^{(4)}(p)$. 
This is the scale that controls the convergence of the expansion of our EFT.
Here, we cannot use $a_0$ for this because it depends 
on our (somewhat \textit{ad hoc}) choice of renormalisation scheme. Also, as 
just noted, the coefficient $a_2$ is not accurately determined and so we are 
left with only $a_1$, the coefficient of the term linear in energy. 

If our theory has a ``natural" scale dependence, $a_1$ should be of the form
$a_1=\hat a_1/\Lambda_0^6$, where $\Lambda_0$ is the scale of the underlying 
physics and $\hat a_1$ is a dimensionless number of order unity. 
The higher-order pion-exchange forces 
that have not been subtracted out can contribute to this coefficient but, since
these are at least of order $Q^4$, their contributions are suppressed by 
$(m_\pi/\Lambda_0)^2$. If we set $\hat a_1=-1$, then our value of $a_1$ implies
a scale of approximately 260~MeV. This is similar to the breakdown scale that 
was recently estimated for the $^1S_0$ wave \cite{birse10}. Such low values for 
the scales suggest the presence of additional low-energy physics that has not 
been considered in the present EFT. One important possible example of this is 
the $\Delta$-resonance. An extended theory, including the $\Delta$ as an explicit
degree of freedom as in Ref.~\cite{kem}, may lead to more natural short-range 
forces.

\section*{Acknowledgments}

KLI is supported by a studentship from the UK STFC. KH was supported by an EU
Free Mover scholarship. MCB acknowledges support from the UK STFC under 
grants PP/F000448/1 and ST/F012047/1. Both KLI and MCB are grateful to the INT 
for its hospitality and to the US Department of Energy for partial support.

\end{document}